\documentclass[12pt,letterpaper]{article}
\usepackage[margin=1in]{geometry}
\usepackage{amsmath,amssymb}
\usepackage{graphicx}
\usepackage{booktabs,tabularx}
\usepackage{subcaption}
\usepackage{xcolor}
\usepackage{enumitem}
\usepackage[section]{placeins}
\usepackage[authoryear,round]{natbib}
\usepackage[hidelinks]{hyperref}
\title{Agentic AI in Automated Markets}
\author{\parbox{0.95\textwidth}{\centering Irene Aldridge\thanks{Contact: \texttt{irene.aldridge@gmail.com}.}, Lourenco Belleau-Bhowmik,
James Chan, Josh Chang, Kelly Deng, Pranav Dhingra, Xinda Ding, Haodong Fang, Satya Hanumanchi,
Carter He, Shuang He, Yadong Hou, Vallab Karanam, Rithvik Kasarla, Tishya Khanna, Jihyun Kim, Sarah
Konety, Abhinav Kotta, Anderson Lo, Jake Luther, Omar Mian, Ishan Patwardhan, Matteo Perona, Chenyue
Qi, Shashank Ramachandran, Shubham Rateria, Jake Swain, David Thai, Michael Twersky, Susan Wu, Dean Xu,
Helenna Yin, Jiachun Zheng$^{1,2}$\\[6pt]
$^1$Cornell Tech}}
\date{}

\begin{document}
\maketitle

\begin{abstract}
Automated markets increasingly host autonomous, learning trading agents and are powered by formulaic
automated market makers (AMMs). We study how heterogeneous agent behavior shapes stress, risk-adjusted
performance, liquidity, and volatility in these types of markets. Three behavioral archetypes are
modeled as agentic reinforcement learning (RL) systems: a rational arbitrageur, a manipulator, and a
retail herd-follower. Each archetype is defined by its reward function within a common, asset-agnostic
exposure-control environment, and each is trained with a value-based Deep Q-Network (DQN) and with
REINFORCE using a learned baseline.

A pilot on hourly cryptocurrency data (AAVE, ETC, SUI) motivates the hypotheses. Herd-followers
produce the most frequent but shortest stress episodes, manipulators the fewest but most persistent,
and rational agents the most stable rolling Sharpe ratios despite higher return volatility. SHAP
attributions tie these patterns to drawdown and cross-asset signals, short-term volatility and
position state, and momentum, respectively. The results motivate behavior-specific regulation rather
than uniform limits on volatility or turnover.

\end{abstract}

\section{Introduction}

Modern markets host a diverse population of traders, a growing share of whom are automated and
increasingly autonomous. In addition, the markets themselves are increasingly automated: an
increasing proportion of market matching happens in markets where market making is automated,
powered by functional Automated Market Makers (AMMs).

Quantitative arbitrageurs, retail investors reacting to news and social media, and actors engaged in
spoofing, wash trading, or front-running all transact in the same venues. Each type leaves a different
footprint on liquidity, price efficiency, and the frequency and persistence of stress. As agentic AI
systems take on more of these roles, proportionate regulation requires knowing which objectives
produce stabilizing behavior and which produce destabilizing behavior.

These questions are hard to answer from aggregate data because behavior types are not labeled in
public datasets. We address this with an agent-based approach. Rather than hand-coding trading rules,
we specify each archetype through its objective and let a reinforcement learning agent discover a
policy that maximizes it. Behavioral assumptions are explicit (they live in the reward function), while
strategy emerges from data. The learned policies then serve as measurement devices: applied to
observable state variables, they yield a stock-day estimate of how intensely each archetype would be
trading.

The framework is asset-agnostic. Any market with a price series, a volume series, and a liquidity proxy
supports it. A pilot on hourly cryptocurrency data, a lightly regulated market with a large retail
share, provides evidence and motivates our hypotheses.

This paper makes five contributions:
\begin{enumerate}
\item We formulate trading as a discrete exposure-control Markov decision process with risk-adjusted,
cost-aware rewards and market-stress penalties, defined over features available in any market.
\item We define three trader archetypes through reward shaping: rational arbitrageur, manipulator, and
retail herd-follower. We explicitly assume that the markets are moderated by Automated Market Makers,
which are deterministic and not agentic at present.
\item We compare a value-based method (DQN) with a policy-gradient method (REINFORCE with a baseline) on
each archetype and use SHAP to attribute each learned policy to its inputs.
\item We translate pilot evidence and testable hypotheses into regulatory recommendations.
\end{enumerate}

Section~\ref{sec:related} reviews related work. Section~\ref{sec:data} describes the data and
features. Section~\ref{sec:method} sets out the environment, algorithms, and archetypes.
Section~\ref{sec:results} reports pilot results and the hypotheses they support.
Sections~\ref{sec:discussion} and~\ref{sec:limits} discuss regulatory implications and limitations.

\section{Related Work}\label{sec:related}

\paragraph{Heterogeneous agents and market quality.} Agent-based computational finance shows that
interactions among fundamentalists, chartists, and noise traders can generate fat tails, volatility
clustering, and crashes \citep{lebaron2006}. Microstructure theory explains how informed traders and
liquidity providers set prices and spreads \citep{kyle1985,glosten1985}. We keep the heterogeneity but
let behavior be learned rather than imposed.

\paragraph{Reinforcement learning for trading.} Direct reinforcement for trading dates to
\citet{moody2001}, with deep variants following \citep{deng2016}. We build on DQN \citep{mnih2015} and
REINFORCE \citep{williams1992}, the canonical value-based and policy-gradient methods, to compare how
algorithm choice interacts with trader objectives.

\paragraph{Retail behavior and manipulation.} Retail investors chase attention and recent performance
\citep{barber2008}. Evidence of manipulation and wash trading exists in both equities and crypto
\citep{gandal2018,cong2023}. We model these behaviors through reward incentives for momentum alignment
and turnover.

\paragraph{Interpretability.} SHAP \citep{lundberg2017} gives additive feature attributions for neural
policies. We use it to verify that each archetype relies on the signals its objective implies.

\section{Data and Features}\label{sec:data}

\paragraph{Sample.} The pilot uses hourly bars for three cryptocurrencies quoted in U.S.\ dollars: AAVE,
the governance token of a decentralized lending protocol; ETC (Ethereum Classic), a proof-of-work
layer-one blockchain; and SUI, a newer layer-one blockchain. The bars come from Yahoo Finance (tickers
\texttt{AAVE-USD}, \texttt{ETC-USD}, and \texttt{SUI20947-USD}) and record the open, high, low, close,
and volume of each UTC hour from 00:00 on March~6, 2024 to 23:00 on February~23, 2026. This window
contains $17{,}280$ calendar hours. Each series has $17{,}245$ or $17{,}246$ bars: $34$--$35$ hours are
missing, almost all in two outages, on October~28, 2025 (11~hours) and November~28--29, 2025
(20~hours), that affect all three assets. There are no duplicate timestamps, no missing fields, and no
bars in which the high and low fail to bracket the open and close. We compute returns only between
consecutive clock hours, so no return spans a gap. The evaluation windows shown in Figures~\ref{fig:crisis}--\ref{fig:vol} span about $1{,}750$ hours.

\paragraph{Volume and the liquidity proxy.} About half of all bars ($50.8\%$ to $51.1\%$, depending on
the asset) report zero volume, at the same rate in every month and every hour of the day. These are not
hours without trading: no bar in the sample has equal high and low prices, and prices move in
zero-volume hours, though less than in other hours (mean absolute return of $0.43$--$0.59\%$, against
$0.68$--$1.02\%$). We therefore treat the volume field as unreliable and use the hourly high--low range,
$\log(\text{High}_t/\text{Low}_t)$, as the liquidity proxy that the framework requires.

\begin{table}[!htbp]
\centering
\small
\begin{tabular}{@{}lrrr@{}}
\toprule
 & AAVE & ETC & SUI\\
\midrule
\multicolumn{4}{@{}l}{\emph{A. Coverage}}\\
Bars observed (of $17{,}280$ calendar hours) & $17{,}246$ & $17{,}245$ & $17{,}246$\\
Missing hours & $34$ & $35$ & $34$\\
Bars reporting zero volume (\%) & $50.8$ & $50.9$ & $51.1$\\
\addlinespace
\multicolumn{4}{@{}l}{\emph{B. Prices (USD)}}\\
First close / last close & $105.21$ / $114.66$ & $33.96$ / $8.35$ & $1.398$ / $0.879$\\
Minimum / maximum close & $73.70$ / $395.13$ & $7.92$ / $39.67$ & $0.473$ / $5.344$\\
Cumulative log return (\%) & $8.6$ & $-140.3$ & $-46.4$\\
\addlinespace
\multicolumn{4}{@{}l}{\emph{C. Hourly log returns (\%)}}\\
Observations & $17{,}240$ & $17{,}239$ & $17{,}241$\\
Mean & $0.001$ & $-0.008$ & $-0.003$\\
Standard deviation & $1.06$ & $0.88$ & $1.17$\\
Annualized volatility & $99.5$ & $82.2$ & $109.9$\\
Skewness & $-0.21$ & $-3.12$ & $-0.36$\\
Excess kurtosis & $11.1$ & $119.9$ & $18.6$\\
Minimum / 1st percentile & $-15.42$ / $-2.95$ & $-32.46$ / $-2.40$ & $-24.90$ / $-3.06$\\
99th percentile / maximum & $2.97$ / $10.18$ & $2.36$ / $9.35$ & $3.35$ / $12.48$\\
Mean high--low range & $1.13$ & $0.88$ & $1.18$\\
\addlinespace
\multicolumn{4}{@{}l}{\emph{D. Stress and serial dependence}}\\
Hours with return below $-2\%$ (\%) & $2.82$ & $1.62$ & $3.51$\\
Stress episodes & $433$ & $255$ & $538$\\
Mean / maximum episode length (hours) & $1.12$ / $3$ & $1.09$ / $4$ & $1.13$ / $4$\\
\quad Mean length if hours were independent & $1.03$ & $1.02$ & $1.04$\\
Variance ratio $\mathrm{VR}(2)$ & $1.017$ & $1.017$ & $1.010$\\
\quad Robust $z$-statistic for $\mathrm{VR}(2)=1$ & $1.28$ & $0.78$ & $0.78$\\
First-order autocorrelation of $|$returns$|$ & $0.241$ & $0.268$ & $0.221$\\
\bottomrule
\end{tabular}
\caption{Summary statistics for the hourly data, March~6, 2024 to February~23, 2026. A stress episode
is a maximal run of consecutive hours with return below $-2\%$; the independence benchmark is
$1/(1-p)$, where $p$ is the share of such hours. $\mathrm{VR}(2)$ is the two-hour variance ratio, and
the $z$-statistic is the heteroskedasticity-robust statistic of \citet{lo1988}.}
\label{tab:data}
\end{table}

\begin{figure}[!htbp]
\centering
\includegraphics[width=\textwidth]{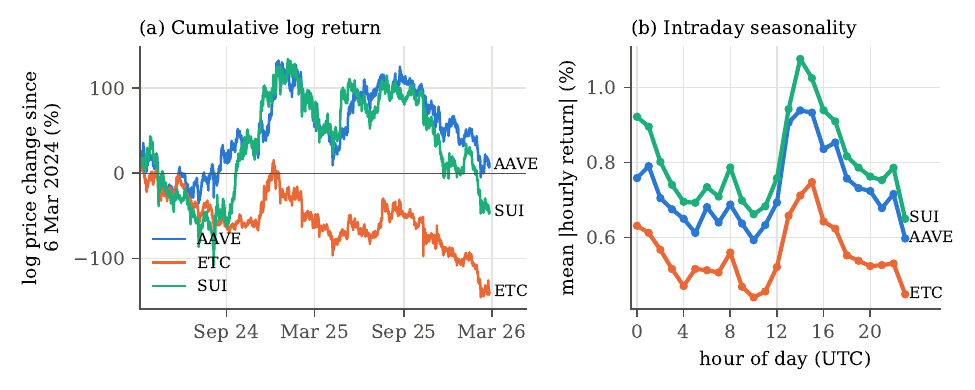}
\caption{\textbf{(a)} Cumulative hourly log return of each asset since March~6, 2024.
\textbf{(b)} Mean absolute hourly return by hour of day (UTC). Volatility peaks between 13:00 and
15:00 UTC, around the U.S.\ equity open, at $1.6$--$1.7$ times its trough.}
\label{fig:data}
\end{figure}

\paragraph{Returns.} Table~\ref{tab:data} summarizes the data. Hourly volatility is high, $82$--$110\%$
annualized, and tails are heavy, with excess kurtosis between $11$ and $120$. All three assets had their
largest one-hour loss in the same hour, 21:00 UTC on October~10, 2025, during a market-wide crash
($-15.4\%$ for AAVE, $-24.9\%$ for SUI, and $-32.5\%$ for ETC, which accounts for most of ETC's
kurtosis). Hourly returns are strongly correlated across assets ($0.64$ for AAVE--SUI, $0.67$ for
ETC--SUI, and $0.73$ for AAVE--ETC), and in $18\%$ of the hours in which any of the assets falls by more
than $2\%$, all three do. Over the sample ETC lost three quarters of its value and SUI more than a
third, while AAVE ended slightly higher after more than tripling and then falling back
(Figure~\ref{fig:data}a).

\paragraph{Stress and serial dependence.} Panel~D describes the market-level counterparts of the
strategy-level metrics in Section~\ref{sec:design}. Hours with returns below $-2\%$, the threshold used
for crisis frequency, make up $1.6$--$3.5\%$ of the sample. They cluster: stress episodes last
$1.09$--$1.13$ hours on average, longer than the $1.02$--$1.04$ hours expected if stress hours were
independent, and absolute returns are strongly autocorrelated ($0.22$--$0.27$). Returns themselves are
nearly serially uncorrelated. The two-hour variance ratio, the efficiency measure in hypothesis H3,
exceeds one by only $0.010$--$0.017$, and the deviation is not statistically significant. Over the full
sample, then, these markets are close to efficient at the hourly horizon, and the persistence of stress
comes from volatility clustering rather than from predictable returns. Tests of H1--H4 should therefore
compare archetype intensity with variation in stress persistence and $|\mathrm{VR}-1|$ over time and
across assets, after controlling for volatility, rather than with their full-sample levels.

\paragraph{Features.} The state $s_t$ that each agent observes is built from prices alone.
Table~\ref{tab:features} reports summary statistics for each price-based feature:
\begin{itemize}[leftmargin=1.2em]
\item \emph{Momentum}, $\log(P_t/P_{t-k})$ for $k=4$ and $k=24$ hours, the signal behind
$\mathrm{mom}_t$ in~\eqref{eq:herd};
\item \emph{Realized volatility}, the standard deviation of the last 24 hourly log returns, which also
serves as the trailing volatility $\hat\sigma_t$ in~\eqref{eq:base};
\item \emph{Drawdown from the 7-day high}, $P_t/\max_{0\le j<168}P_{t-j}-1$;
\item \emph{Cross-asset return}, the equal-weighted mean of the other two assets' hourly log returns;
\item \emph{High--low range}, $\log(\text{High}_t/\text{Low}_t)$, the liquidity proxy;
\item \emph{Hour of day} (UTC); and
\item the agent's \emph{current position} $e_{t-1}$, which belongs to the agent's state rather than to
the data.
\end{itemize}

\begin{table}[!htbp]
\centering
\small
\begin{tabular}{@{}lrrrrr@{}}
\toprule
Feature (\%) & Mean & SD & 5th pct. & Median & 95th pct.\\
\midrule
\multicolumn{6}{@{}l}{\emph{AAVE}}\\
Momentum, 4 hours & $0.00$ & $2.14$ & $-3.23$ & $-0.03$ & $3.34$\\
Momentum, 24 hours & $0.00$ & $5.21$ & $-7.75$ & $-0.16$ & $8.26$\\
Realized volatility, 24 hours & $0.96$ & $0.46$ & $0.45$ & $0.85$ & $1.75$\\
Drawdown from 7-day high & $-8.71$ & $6.95$ & $-22.43$ & $-7.27$ & $-0.25$\\
Cross-asset return, 1 hour & $-0.01$ & $0.94$ & $-1.38$ & $-0.01$ & $1.39$\\
High--low range, 1 hour & $1.13$ & $1.05$ & $0.33$ & $0.90$ & $2.66$\\
\addlinespace
\multicolumn{6}{@{}l}{\emph{ETC}}\\
Momentum, 4 hours & $-0.03$ & $1.75$ & $-2.58$ & $-0.02$ & $2.42$\\
Momentum, 24 hours & $-0.20$ & $4.18$ & $-6.58$ & $-0.17$ & $6.00$\\
Realized volatility, 24 hours & $0.75$ & $0.46$ & $0.33$ & $0.64$ & $1.51$\\
Drawdown from 7-day high & $-7.39$ & $5.68$ & $-18.16$ & $-6.25$ & $-0.31$\\
Cross-asset return, 1 hour & $0.00$ & $1.02$ & $-1.53$ & $0.00$ & $1.52$\\
High--low range, 1 hour & $0.88$ & $0.85$ & $0.24$ & $0.67$ & $2.17$\\
\addlinespace
\multicolumn{6}{@{}l}{\emph{SUI}}\\
Momentum, 4 hours & $-0.01$ & $2.37$ & $-3.59$ & $-0.05$ & $3.69$\\
Momentum, 24 hours & $-0.07$ & $5.86$ & $-8.36$ & $-0.39$ & $10.04$\\
Realized volatility, 24 hours & $1.07$ & $0.50$ & $0.49$ & $0.97$ & $1.91$\\
Drawdown from 7-day high & $-9.89$ & $6.91$ & $-22.66$ & $-8.87$ & $-0.51$\\
Cross-asset return, 1 hour & $0.00$ & $0.91$ & $-1.32$ & $0.00$ & $1.31$\\
High--low range, 1 hour & $1.18$ & $1.02$ & $0.36$ & $0.95$ & $2.78$\\
\bottomrule
\end{tabular}
\caption{Summary statistics for the price-based features, in percent, over the full sample. The first
167 hours are dropped so that every rolling window is complete.}
\label{tab:features}
\end{table}

Three features connect directly to the statistics above. Time of day matters because volatility has a
pronounced intraday cycle (Figure~\ref{fig:data}b), which is consistent with its appearance among the
top three SHAP features of every archetype (Section~\ref{sec:shap}). The cross-asset return carries
information because the three assets move together and crash together. And the drawdown signal
varies widely: the typical drawdown from the 7-day high is $6$--$9\%$, and one hour in twenty sits
$18$--$23\%$ or more below it.

\paragraph{Stress regime.} The stress indicator in~\eqref{eq:base} flags hours in a market crash regime. For the CRSP tests it flags days on which
VWRETD is below its 5th percentile or the VIX jumps above its 95th percentile (Section~\ref{sec:env}).

\section{Methodology}\label{sec:method}

\subsection{Environment}\label{sec:env}

Trading is a discrete exposure-control problem. At each step $t$ the agent observes state $s_t$ and
chooses an action $a_t\in\{\text{sell},\text{hold},\text{buy}\}$, which moves exposure $e_t$ by $-0.5$,
$0$, or $+0.5$ within $[0,1]$. Exposure $0$ is flat, $0.5$ is half invested, $1$ is fully long. The base
reward is a volatility-normalized net return:
\begin{equation}\label{eq:base}
r^{\mathrm{base}}_t=\frac{e_tR_{t+1}-c\,|e_t-e_{t-1}|}{\hat\sigma_t+\epsilon}
-\lambda_{\mathrm{stress}}\,\mathbf 1[\mathrm{stress}_t]\,e_t .
\end{equation}
Here $R_{t+1}$ is the next-period asset return, $c$ is the proportional transaction cost,
$\hat\sigma_t$ is trailing volatility, and the stress indicator flags market crash regimes (in CRSP:
VWRETD below its 5th percentile or a VIX jump above its 95th percentile).

\subsection{Algorithms}

\paragraph{DQN (value-based).} A two-layer multilayer perceptron estimates $Q(s,a)$ for the three
actions. A replay buffer breaks the temporal correlation between updates, and a target network
synchronized every $N$ steps stabilizes bootstrapped targets. Exploration is $\varepsilon$-greedy with
linear decay.

\paragraph{REINFORCE with baseline (policy gradient).} A two-layer policy network outputs a softmax
over the three actions. A separate two-layer value network estimates the expected discounted return and
serves as a baseline. The advantage is the Monte Carlo return minus the baseline estimate, which reduces
gradient variance without introducing bias. Returns are standardized within each batch to stabilize
training. Because the critic is used only as a baseline and not for bootstrapping, the method is
REINFORCE with a baseline rather than a full actor-critic.

\subsection{Trader archetypes}

Each archetype modifies the base reward~\eqref{eq:base}. Archetypes are trained in separate
environments, so behavioral effects are isolated; strategic interaction is not modeled
(Section~\ref{sec:limits}). Unlike traditional markets, our environment is presumed to be matched by
Automated Market Makers (AMMs), which at present are deterministic and functional. We, therefore,
exclude market makers from this research.

\paragraph{Rational arbitrageur.} Maximizes risk-adjusted return with additional penalties on realized
volatility and turnover. It acts conservatively and reduces exposure as risk rises.

\paragraph{Manipulator.} A stylized proxy for front-running and wash trading. It receives a bonus
proportional to turnover, $\kappa_{\mathrm{turn}}|e_t-e_{t-1}|$, and a short discount horizon, which
favors rapid position cycling and short-term gains.

\paragraph{Retail herd-follower.} Mimics trend-chasing and fear-of-missing-out behavior. It receives a
bonus for aligning exposure with recent momentum,
\begin{equation}\label{eq:herd}
\kappa_{\mathrm{mom}}\,\mathrm{sign}(\mathrm{mom}_t)\,(e_t-0.5).
\end{equation}

\subsection{Hyperparameters}

Reward weights $(\kappa_{\mathrm{turn}},\kappa_{\mathrm{mom}},\lambda_{\mathrm{stress}})$ are fixed
across assets and set for the validation window. Each configuration is trained with five random seeds.
The reported results average across those seeds.

\section{Experimental Design}\label{sec:design}

We trained three archetypes of agents and evaluated the performance of each strategy.

\subsection{Strategy-level metrics}

These metrics (Table~\ref{tab:metrics}) describe the strategies, not the market. Because agents are
price-takers in Stage~1, a strategy's Sharpe ratio measures how well it exploits the market, not how
efficient the market is. Stage~2 addresses market-level questions.

\begin{table}[!htbp]
\centering
\begin{tabular}{@{}ll@{}}
\toprule
Metric & Definition\\
\midrule
Crisis frequency & Hours in a 24-hour window with return below $-2\%$\\
Crisis persistence & Visual run length\\
Efficiency & Rolling 7-day Sharpe\\
Return volatility & Rolling 24-hour $\sigma$\\
\bottomrule
\end{tabular}
\caption{Stage~1 strategy-level metrics}
\label{tab:metrics}
\end{table}

\subsection{Hypotheses}

\begin{table}[!htbp]
\centering
\small
\begin{tabularx}{\textwidth}{@{}lXp{3.6cm}l@{}}
\toprule
ID & Hypothesis & Expected sign & Source\\
\midrule
H1 & Retail intensity raises crash frequency but not persistence & $\beta_{\mathrm{RET}}>0$ on frequency
& Pilot\\
H2 & Manipulator intensity raises crash persistence & $\beta_{\mathrm{MAN}}>0$ on persistence & Pilot\\
H3 & Rational intensity lowers inefficiency (variance-ratio deviation) & $\beta_{\mathrm{RA}}<0$ on
$|\mathrm{VR}-1|$ & Pilot\\
H4 & Rational intensity raises volatility without raising crash frequency & $\beta_{\mathrm{RA}}>0$ on
volatility, $\approx0$ on frequency & Pilot\\
\bottomrule
\end{tabularx}
\caption{Testable hypotheses}
\label{tab:hyp}
\end{table}

\section{Results}\label{sec:results}

This section reports the cryptocurrency pilot for the rational, manipulator, and retail archetypes
(Sections~\ref{sec:algo}--\ref{sec:sharpe}). Pilot findings are based on one training run per
configuration and visual inspection of rolling metrics, so they are motivating evidence rather than
formal tests.

\subsection{Algorithm comparison}\label{sec:algo}

DQN explores through $\varepsilon$-greedy action selection, which produces abrupt switches between
policies during training. REINFORCE samples from a smooth action distribution, which yields gentler
exploration and, in most configurations, less erratic turnover. DQN learned faster per sample because
its replay buffer reuses past transitions. REINFORCE was more prone to policy collapse: on AAVE and SUI
the rational REINFORCE agent converged to zero exposure, producing a flat zero line in both the Sharpe
and volatility figures. We treat that as a degenerate solution under a strong volatility penalty, not
as evidence of rational behavior, and it motivates the multi-seed protocol.

\subsection{Feature attributions}\label{sec:shap}

We applied SHAP DeepExplainer to the REINFORCE action probabilities and the DQN Q-values, using
randomly sampled historical states as the background distribution and the test period as the
evaluation set. Mean absolute SHAP values identify the three most influential features per archetype
(Table~\ref{tab:shap}).

Time of day ranks in the top three for every archetype, which likely reflects intraday liquidity and
volatility seasonality in crypto.

\begin{table}[!htbp]
\centering
\small
\begin{tabularx}{\textwidth}{@{}p{2.6cm}p{3.8cm}X@{}}
\toprule
Archetype & Top-3 features (pilot) & Interpretation\\
\midrule
Rational arbitrageur & 7-day drawdown, cross-asset return, time of day & Cuts exposure in drawdowns;
uses cross-market information\\
Manipulator & Short-term volatility, time of day, current position & Cycles positions around short-term
volatility\\
Retail herd-follower & 24-hour momentum, 4-hour momentum, time of day & Momentum dominates; amplifies
trends and suffers at reversals\\
\bottomrule
\end{tabularx}
\caption{Top SHAP features by archetype}
\label{tab:shap}
\end{table}

\subsection{Crisis frequency and persistence}

Retail agents show the most frequent stress episodes, but these are short, consistent with shallow
drawdowns from momentum overreaction (Figure~\ref{fig:crisis}). Rational agents show a moderate
frequency, with controlled exposure limiting both occurrence and duration. Manipulators show the fewest
episodes, but stress persists the longest when it occurs, most visibly around the major crash regimes
near hours 1,450--1,500. All strategies have far fewer stress hours than the market's own crash regime
because none are fully invested at all times.

\subsection{Risk-adjusted performance and volatility}\label{sec:sharpe}

Rolling 7-day Sharpe ratios swing widely for every agent, roughly between $-20$ and $+20$ annualized
(Figure~\ref{fig:sharpe}). Rational agents are most stable where they trade. Manipulators show spikes
under favorable conditions, followed by sharp deteriorations. Retail agents are weakest during
reversals; for example, on ETC between hours 300 and 800.

Return volatility tells a different story from stress (Figure~\ref{fig:vol}). The active rational
agents (DQN on AAVE, REINFORCE on ETC) have the highest baseline volatility, reflecting continual
exposure adjustments. Manipulators have lower baseline volatility, punctuated by sharp spikes during
stress periods, up to about 0.02 on SUI. Retail volatility sits in between. Higher volatility,
therefore, does not imply more stress: the most volatile strategy is not the one with the most
persistent drawdowns.

\newcommand{\threepanel}[1]{%
\begin{subfigure}{0.66\textwidth}\centering\includegraphics[width=\linewidth]{figs/#1_aave.png}\caption{AAVE}\end{subfigure}\\[2pt]
\begin{subfigure}{0.66\textwidth}\centering\includegraphics[width=\linewidth]{figs/#1_etc.png}\caption{ETC}\end{subfigure}\\[2pt]
\begin{subfigure}{0.66\textwidth}\centering\includegraphics[width=\linewidth]{figs/#1_sui.png}\caption{SUI}\end{subfigure}}

\begin{figure}[p]
\centering
\threepanel{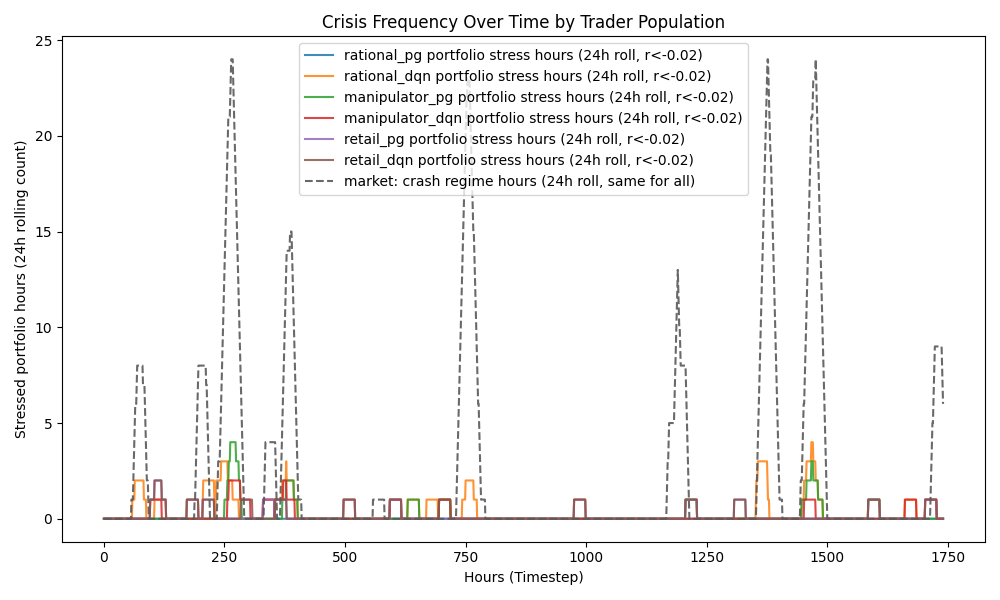}
\caption{Hours with net strategy return below $-2\%$ in a rolling 24-hour window, by agent. Dashed line:
hours the market is in a crash regime.}
\label{fig:crisis}
\end{figure}

\begin{figure}[p]
\centering
\threepanel{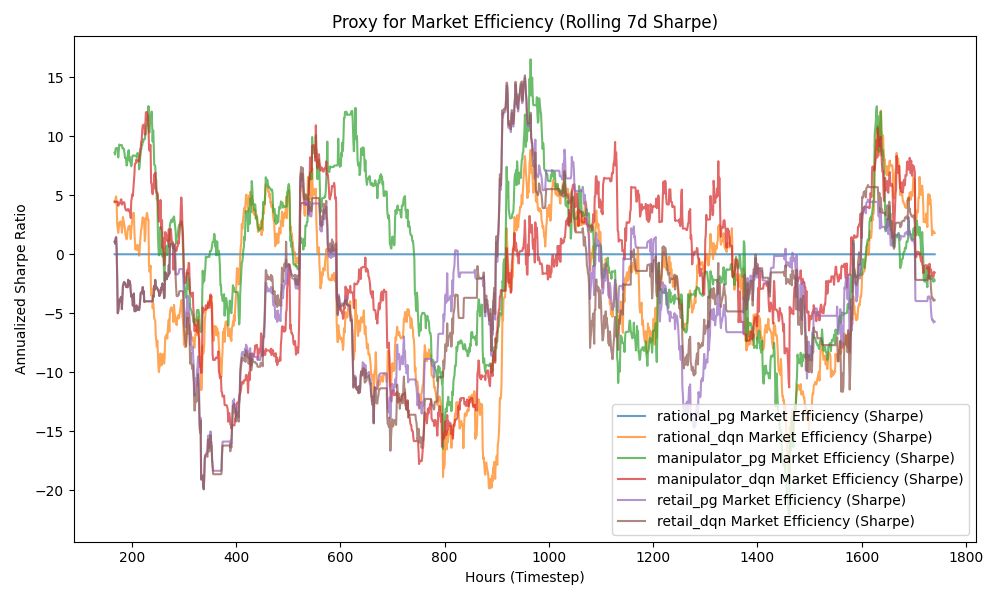}
\caption{Rolling 7-day annualized Sharpe ratio of net strategy returns, by agent.}
\label{fig:sharpe}
\end{figure}

\begin{figure}[p]
\centering
\threepanel{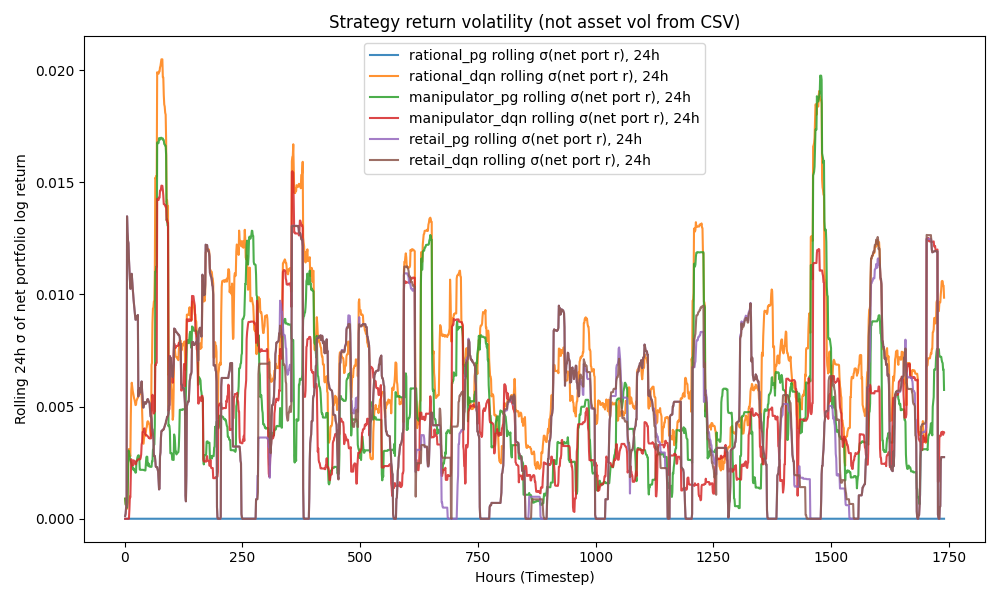}
\caption{Rolling 24-hour standard deviation of net strategy log returns, by agent.}
\label{fig:vol}
\end{figure}

\section{Discussion and Regulatory Implications}\label{sec:discussion}

The pilot suggests that the regulation of automated and agentic trading should target behaviors rather
than apply uniform limits on volatility or turnover. Each recommendation below is tied to a hypothesis
that the CRSP tests can confirm or reject.

\paragraph{Herding (H1).} If retail intensity predicts frequent short stress episodes, the case
strengthens for investor-protection tools that slow pro-cyclical flows: risk disclosure at the point of
trade, leverage and margin limits for retail accounts, and single-stock circuit breakers such as the
U.S.\ limit-up/limit-down mechanism. Crypto venues largely lack these tools, so the pilot setting is
where the gap is widest.

\paragraph{High-turnover and manipulative strategies (H2).} If manipulator intensity predicts
persistent stress, surveillance should prioritize abnormal turnover and order-cycling patterns, and
venues could consider fees that scale with order-to-trade or cancellation ratios. These tools target
activity that creates the appearance of liquidity without its substance.

\paragraph{Rational risk management (H3, H4).} Rational agents produced the highest return volatility in
the pilot without the most persistent stress. Volatility alone is therefore a poor regulatory trigger.
Rules that penalize all volatility may suppress the adaptive repositioning that keeps prices
informative.

Overall, effective policy should distinguish between stabilizing and destabilizing behavior. The
framework proposed here offers a way to measure that distinction using public data.

\section{Limitations and Future Work}\label{sec:limits}

\begin{itemize}
\item \textbf{No price impact or interaction.} Agents are price-takers in separate environments, so
Stage~1 cannot show how one archetype's trades move prices or affect another's. Stage~2 links archetype
intensity to market outcomes, but it is correlational. A shared limit-order-book simulator with all
three archetypes trading together is the natural extension.
\item \textbf{Policies as signal composites.} Archetype intensity is a nonlinear function of public
signals. Controls for size, value, momentum, volatility, and turnover reduce, but do not eliminate, the
risk that intensity proxies for a known characteristic. The placebo and external-validation tests
address this.
\item \textbf{Coarse action space.} Exposure moves in steps of 0.5 within $[0,1]$, with no short
positions. A finer or continuous action space and short selling would allow for richer behavior.
\item \textbf{Pilot scale.} The pilot used one run per configuration on about 1,750 hours per asset.
Policy collapse in the rational REINFORCE agent shows that multi-seed training and reward-weight tuning
are needed before drawing formal conclusions.
\end{itemize}

\section{Conclusion}

We model three trader archetypes as agentic reinforcement learning systems whose behavior follows from
their objectives, and we propose using the trained policies to measure archetype intensity in public
market data. The cryptocurrency pilot shows distinct footprints: frequent, shallow stress for
herd-followers; rare, persistent stress for manipulators; and volatile but stable risk-adjusted
performance for rational agents. The findings support regulation that targets herding and
manipulation while preserving adaptive risk-taking.

\bibliographystyle{plainnat}
\bibliography{references}

\end{document}